\documentclass{elsart}
\usepackage{graphics}
\begin{document}
%\runauthor{Albert-L\'aszl\'o Barab\'asi}
\begin{frontmatter}
\title{Mean-field theory for scale-free random networks}
\author{Albert-L\'aszl\'o Barab\'asi\thanksref{Someone},}
\author{R\'eka Albert,}
\author{Hawoong Jeong}
 
\address{Department of
Physics, University of Notre-Dame, Notre-Dame, IN 46556, USA}
\thanks[Someone]{Tel.: +1 219 631 5767; Fax: +1 219 631 5952; e-mail: alb@nd.edu}
\begin{abstract}
Random networks with complex topology are common in Nature,
describing systems as diverse as the 
world wide web or social and business
networks. Recently,
it has been demonstrated that most large networks for which topological
information is available display scale-free features.
Here we study the scaling properties of the recently
introduced scale-free model, that can account for 
the observed power-law distribution of the
connectivities.
We develop a mean-field method to predict the growth dynamics of the
individual vertices, and use this to calculate analytically
the connectivity distribution and the scaling exponents.
The mean-field method can be used to address 
the properties of two variants of the scale-free
model, that do not display power-law scaling.

\bigskip
\noindent {\it PACS}:
\end{abstract}
\end{frontmatter}

\section{Introduction}
Contemporary science has been particularly successful in addressing the
physical properties of systems that are composed of many identical elements 
interacting through mainly local interactions.
For example, many successes of materials science and solid state
physics are based on the fact that most solids are made of relatively few types of elements that exhibit spatial order by forming 
a crystal lattice. Furthermore, these elements
are coupled by local, nearest neighbor interactions. However,
the inability of contemporary science to describe systems composed of 
non-identical elements that have diverse and nonlocal interactions 
currently limits advances in many disciplines, ranging from molecular 
biology  to computer science \cite{complex}. The 
difficulty in describing these systems lies partly in their 
topology: many of them form complex networks, 
whose vertices are the elements of the system and edges 
represent the interactions between them. For example, living 
systems form a huge 
genetic network, whose vertices are proteins,  
the edges representing the chemical interactions between 
them \cite{biology}.
Similarly, a large network is formed by the nervous 
system, whose vertices are the nerve cells,
connected by axons \cite{neural}. But equally complex 
networks occur in social science, where vertices are individuals,
organizations or countries, and the edges characterize the social interaction 
between them \cite{social}, in the business world, where vertices are 
companies and edges represent diverse business relationships,
or describe the world wide web (www), whose 
vertices are  HTML documents connected by links pointing  from one 
page to another \cite{clever,diam}. 
Due to their large size and the complexity of the 
interactions, the topology of these networks is largely unknown 
or unexplored.

Traditionally, networks of complex topology have been described using 
the random graph theory of Erd\H{o}s and R\'enyi (ER) \cite{random}. 
However, while it has been much investigated in 
combinatorial graph theory,
in the absence of data on large networks the 
predictions of the  ER theory were rarely tested in the real world. 
This is changing very fast lately: driven by the 
computerization of data acquisition, topological information 
on various real world networks is 
increasingly available.
Due to the importance of understanding the topology of 
some of these systems,
it is likely that in the near future we will
witness important advances in this direction.
Furthermore, it is also possible that seemingly random networks in Nature
have rather complex internal structure, that cover generic features,
common to many systems.
 Uncovering the universal properties characterizing the formation and 
the topology of complex networks could bring about the 
much coveted revolution beyond reductionism \cite{complex}.

A major step in the direction of understanding the generic features
of network development was the recent discovery 
of a surprising  degree of self-organization characterizing  
the large scale properties of complex networks. Exploring 
several large databases describing the topology of 
large networks, that span as diverse fields as the www or 
the citation patterns in science, recently Barab\'asi 
and Albert (BA) have demonstrated\cite{ba} that independently 
of the nature of the system and the identity 
of its constituents, the probability 
$P(k)$ that a vertex in the  network is connected to $k$ 
other vertices decays as a power-law, following $P(k) \sim k^{-\gamma}$.  
The generic feature of this observation was supported 
by four real world examples.
In the collaboration graph of movie actors, each actor is represented by a
 vertex, two actors being connected if they were 
casted in the same movie. The probability that 
an actor has $k$ links was found to follow 
a power-law for large $k$, i.e. 
$P(k) \sim k^{-\gamma_{actor}}$, 
where $\gamma_{actor}=2.3\pm 0.1.$ 
A rather complex network with over $300$ 
million vertices \cite{giles} is the  
www,  where a vertex is a document and the edges are 
the links pointing from one document to another. 
The topology 
of this graph determines the web's connectivity 
and, consequently, our effectiveness in locating information 
on the www \cite{clever}. Information about $P(k)$  
can be obtained using robots \cite{diam}, indicating 
that the probability that $k$ documents 
point to a certain webpage follows a power-law,  with 
$\gamma_{www}^{in}=2.1$ \cite{note}, and the probability
that a certain web document contains $k$ outgoing links follows
a similar distribution, with $\gamma_{www}^{out}=2.45$. 
A network whose topology 
reflects the historical patterns of urban and industrial 
development is the electrical powergrid of western US, the  vertices
 representing generators, transformers and substations, the edges  
corresponding to the high voltage transmission lines between 
them \cite{small_world}.
The connectivity distribution is again best approximated with 
a power-law with an exponent $\gamma_{power}\simeq 4$. 
Finally, a rather large, complex network is formed by the 
citation patterns of the scientific 
publications, the vertices standing for papers, 
the edges representing links to the articles 
cited in a paper. Recently Redner \cite{redner} has 
shown that the probability that a paper is cited 
$k$ times (representing the 
connectivity of a paper within the network) follows 
a power-law with exponent $\gamma_{cite}=3$. 
These results offered the first evidence that large networks 
self-organize into a scale-free state, a feature unexpected by all 
existing random network models. To understand the origin 
of this scale invariance, BA have shown that existing 
network models fail to incorporate two key features of real 
networks:  First, networks continuously grow by the addition 
of new vertices, and second, new vertices connect 
preferentially  to highly 
connected vertices. Using a model incorporating these ingredients, 
they demonstrated that the combination of growth and 
preferential attachment is ultimately responsible for the scale-free 
distribution and power-law scaling observed in real networks.

The goal of the present paper is to investigate 
the properties of the scale-free model introduced by BA \cite{ba}, 
aiming to identify its scaling properties and compare them 
with other network models 
intended to describe the large scale properties of random networks.
We present a mean field theory that allows us to predict
the dynamics of individual vertices in the system, and to
calculate analytically the connectivity distribution.
We apply the same method to uncover the scaling
properties of two versions of the BA model, that are missing one of
the ingredients needed to reproduce the power-law scaling.
Finally, we discuss various extensions of the BA model,
that could be useful in addressing the properties of real networks. 

\section{Earlier network models}
\subsection{The Erd\H{o}s-R\'enyi model}
Probably the oldest and most investigated random 
network model has been introduced
by Erd\H{o}s and R\'enyi (ER) \cite{random}, who were the first 
to study the statistical aspects of random graphs by 
probabilistic methods.
In the model we start with $N$ vertices and no bonds (see Fig.$\,$1a).
With probability $p_{\mathrm{ER}}$, we connect each pair of
vertices with a line (bond or edge), generating a random network.
The greatest discovery of 
ER was that many properties of these graphs appear quite 
suddenly, at a threshold value of $p_{\mathrm {ER}}(N)$. A property of 
great importance for the 
topology of the graph is the appearance of 
trees and cycles. A tree of order $k$ is a connected 
graph with $k$ vertices and $k-1$ edges, while a cycle of 
order $k$ is a cyclic sequence of $k$ edges such that every 
two consecutive edges and only 
these have a common vertex. ER have demonstrated that 
if $p_{\mathrm {ER}}\sim c/N$ with $c<1$, then almost all vertices belong to 
isolated trees, but there is an abrupt change at $p_{\mathrm {ER}}\sim 1/N$, (i.e. $c=1$),
when cycles of all orders appear.
In the physical literature the ER model is often referred
to as infinite dimensional percolation, that is known to belong to the
universality class of mean field percolation \cite{stauffer}.
In this context $p_c \sim 1/N$ is the percolation threshold of the system.
For $p<p_c$ the system is broken into many small clusters, while at $p_c$
a large cluster forms, that in the asymptotic limit contains all vertices.

To compare the ER model with other network models,
we need to focus on the connectivity distribution. 
As Erd\H{o}s and R\'enyi have shown in their seminal work,
the probability that a vertex has $k$ edges follows the Poisson 
distribution 
\begin{equation}
P(k)=e^{-\lambda}{\lambda}^k/k! ,
\end{equation}
where  
\begin{equation}
\lambda = \left( \begin{array}{c} N-1 \\ k \end{array} \right) p_{\mathrm {ER}}^k (1-p_{\mathrm {ER}})^{N-1-k} ,
\end{equation}
its expectation value being $(N-1)p_{\mathrm {ER}}$.
For sake of comparison, in Fig.$\,$2a we show $P(k)$ for
different values of $p_{\mathrm {ER}}$.
 
\begin{figure}
\resizebox{\textwidth}{!}{\includegraphics{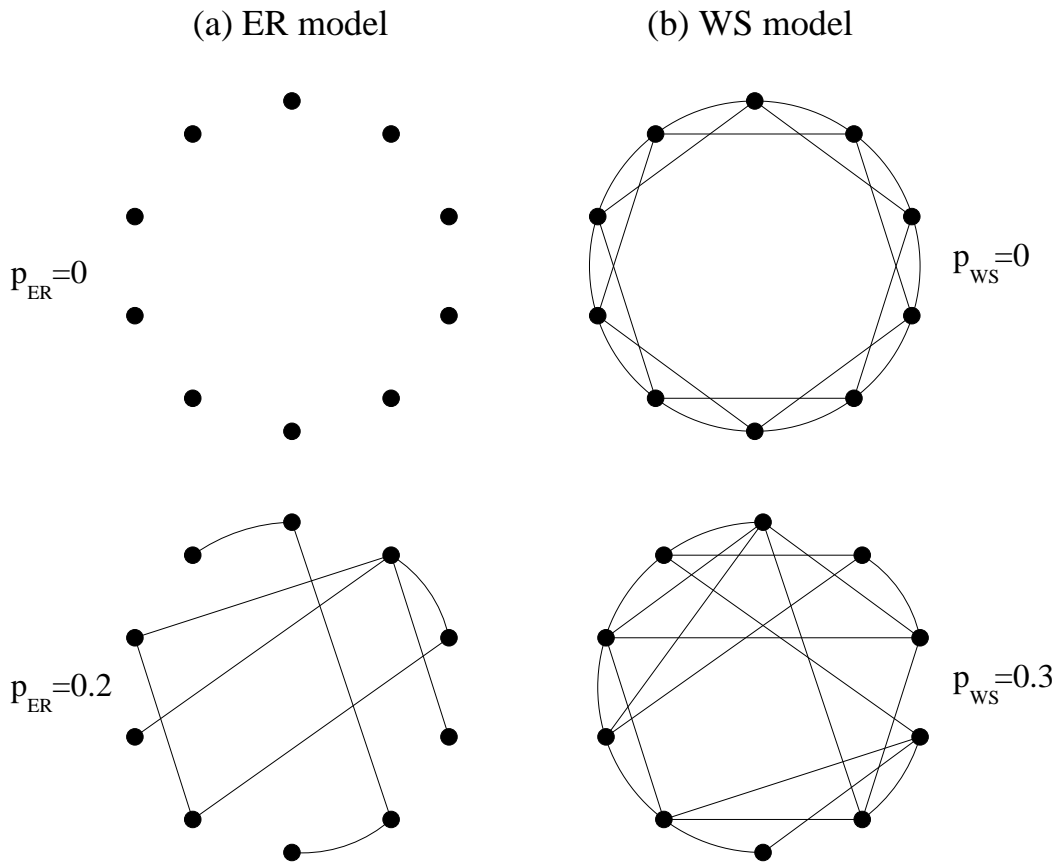}}
\caption{Schematic illustration of the Erd\H{o}s-R\'enyi (ER) and Watts-Strogatz (WS) models. (a) A random network described by the ER model has N vertices connected with probability $p_{\mathrm {ER}}$, the total number of edges in the system being $n=p_{\mathrm {ER}}N(N-1)/2$. The example presents a network of $N=10$ vertices for $p_{\mathrm {ER}}=0$ and $p_{\mathrm {ER}}=0.2$. At $p_{\mathrm {ER}}=0$ there are no edges in the system. We select each pair of vertices and connect them with probability $p_{\mathrm {ER}}=0.2$. The figure shows the result of this process, the network having $n=9$ edges. For $p_{\mathrm {ER}}=1$ the model leads to a fully connected network. (b) The WS model starts with a regular one dimensional lattice with edges between the nearest and next-nearest neighbors, thus the average connectivity is $\langle k\rangle=4$. Then a fraction $p_{\mathrm {WS}}$ of the edges is rewired randomly (their endpoint is changed to a randomly selected vertex). The example presents a network of $N=10$ vertices. For $p_{\mathrm {WS}}=0$ the system is a regular lattice with $2N=20$ edges. For $p_{\mathrm {WS}}=0.3$, $2p_{\mathrm {WS}}N=6$ edges have been rewired to randomly selected vertices. Note that for $p_{\mathrm {WS}}=1$ we obtain a random network, equivalent to that obtained for the ER model with $p_{\mathrm {ER}}=\langle k \rangle/N=0.4$.}
\end{figure}

\subsection{The small-world model}
Aiming to describe the transition from a locally ordered 
system to a random network, recently Watts and Strogatz (WS) have introduced
a new model \cite{small_world}, 
that is often referred to as small-world network. The topological
properties of the network generated by this model have been the subject of much
attention 
lately \cite{amaral,sw1,sw2,sw3,sw4,sw5,sw6,sw7,sw8,sw9,sw10,sw11,sw12}.
The WS model begins with a one-dimensional lattice of $N$ 
vertices with bonds between the nearest and next-nearest 
neighbors (in general, the algoritm can include neighbors up to an order $n$, such that the coordination number of a vertex is $z=2n$) and periodic boundary conditions (see Fig.$\,$1b). Then each 
bond is rewired with probability $p_{\mathrm {WS}}$, where rewiring in this context 
means shifting 
one end of the bond to a new vertex chosen at random from the 
whole system, with the constraint that no two vertices can 
have more than one bond, and no vertex can have a bond with 
itself. For $p_{\mathrm {WS}}=0$ the lattice is highly clustered, and the average 
distance between two vertices $\langle l\rangle$ grows linearly 
with $N$, while for $p_{\mathrm {WS}}=1$ the system  becomes a random graph, 
poorly clustered and $\langle l \rangle$ grows  logarithmically 
with $N$. WS found that in the interval $0<p_{\mathrm {WS}}<0.01$ the model exhibits
 small-world properties\cite{small_world_book}, 
($\langle l\rangle\simeq \langle l\rangle_{random}$), 
while it remains highly clustered.

The connectivity distribution of the WS model depends strongly on $p_{\mathrm {WS}}$: 
for $p_{\mathrm{WS}}=0$ we have $P(k)=\delta(k-z)$, where $z$ is the coordination 
number of the lattice, while for finite $p_{\mathrm{WS}}$, 
$P(k)$ is still peaked around $z$, but it gets broader. 
Ultimately, as $p_{\mathrm{WS}} \rightarrow 1$, the distribution $P(k)$ approaches the 
connectivity distribution of a random graph, i.e. the distribution converges to that obtained for the ER model with $p_{\mathrm {ER}}=z/N$ (see Fig.$\,$2b).

\begin{figure}[ht]
\rotatebox{270}{\resizebox{\textwidth}{!}{\includegraphics{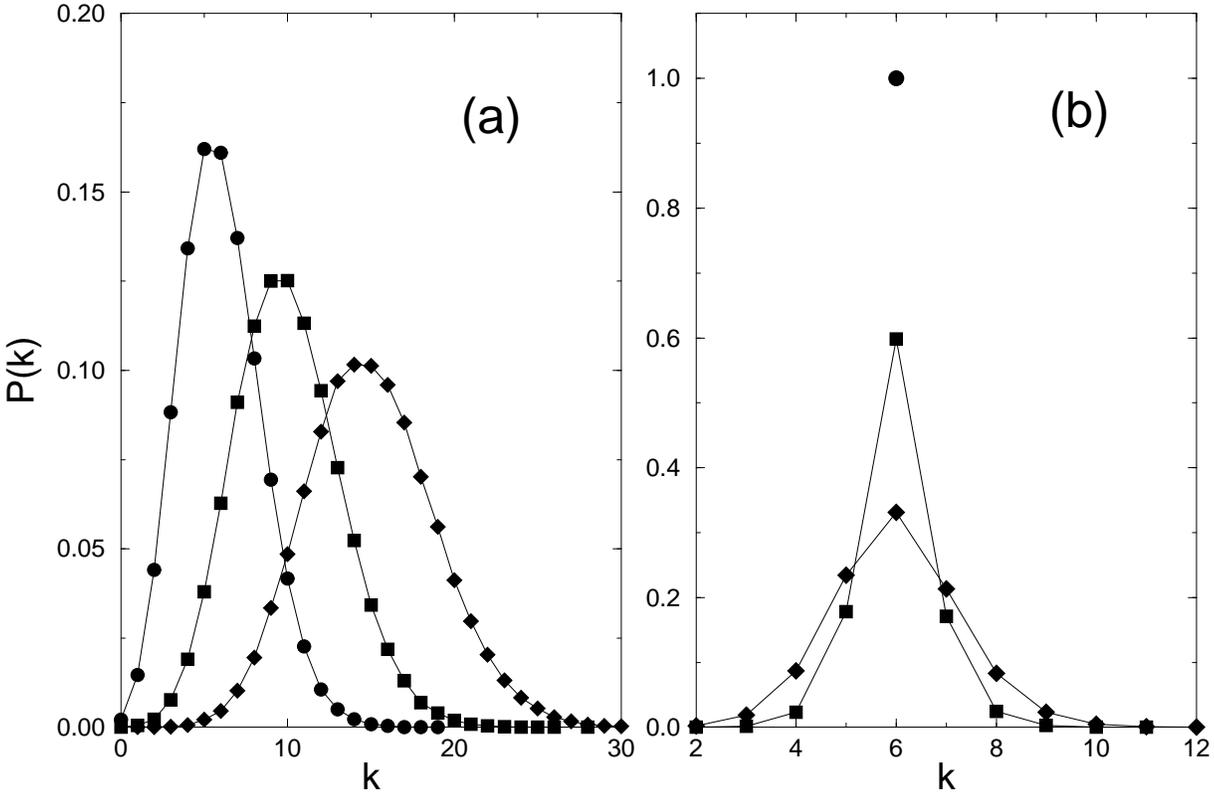}}}
\vspace{-3cm}
\caption{Connectivity distributions for the ER and WS models. (a) $P(k)$ in the ER model for $N=10,000$ and for $p_{\mathrm {ER}}=0.0006$ (circles), $p_{\mathrm {ER}}=0.001$ (squares) and $p_{\mathrm {ER}}=0.0015$ (diamonds). (b) $P(k)$ in the WS model for $N=10,000$, $\langle k\rangle=6$ and three rewiring probabilities $p_{\mathrm {WS}}=0$ (circle, corresponding to the delta-function $\delta(k-6)$), $p_{\mathrm {WS}}=0.1$ (squares) and $p_{\mathrm {WS}}=0.3$ (diamonds). }
\end{figure}

\section{The scale-free model}
A common feature of the models discussed in the previous
section is that they both predict that the probability distribution
of the vertex connectivity, $P(k)$, has an exponential cutoff,
and has a characteristic size $\left< k\right>$,
that depends on $p$. In contrast, as we mentioned in the Introduction,
many systems in nature have 
the common property that $P(k)$ is free of scale, following
a power-law distribution over many orders of magnitude.
To understand the origin of this discrepancy, BA have argued that
there are two generic aspects of real networks that are not 
incorporated in these models \cite{ba}. First, both models assume that we 
start with a fixed number ($N$) of vertices, that are then 
randomly connected (ER model), or reconnected (SW model), without 
modifying $N$. In contrast, most real world networks 
are {\it open}, i.e. they form by the continuous addition of 
new vertices to the system, thus the number of vertices, $N$, 
increases throughout the lifetime of the network. For example, 
the actor network 
grows by the addition of new actors to the system, the 
www grows exponentially in time by the addition of new 
web pages, the research literature constantly 
grows by the publication of new papers. Consequently, a 
common feature of these systems is 
that the {\it network continuously expands by the addition of 
new vertices} that are connected to the vertices already present 
in the system.

Second, the random network models assume that the probability that 
two vertices are connected is random and uniform. In contrast, 
most real networks exhibit {\it preferential connectivity}. For 
example, a new actor is casted most likely in a supporting role,
with more established, well known actors. 
Similarly, a newly created webpage will more likely include 
links to well 
known, popular documents with already high connectivity, or a 
new manuscript is more likely to cite a well known and thus 
much cited paper than its less cited and consequently less 
known peer. These examples indicate that the probability with 
which a new 
vertex connects to the existing vertices is not uniform, but 
there is a {\it higher probability to be linked to a vertex 
that already has a large number of connections}.
The scale-free model introduced by BA, incorporating only 
these two ingredients, naturally leads to the observed 
scale invariant distribution. 
The model is defined in two steps (see Fig.$\,$3):

(1) {\it Growth}:  
Starting with a small number ($m_0$) of vertices, at every timestep we add a
 new vertex with $m$($\leq m_0$) edges (that will be connected
to the vertices already present in the system). 

(2) {\it Preferential attachment}: When choosing the vertices to which
the new vertex connects, we assume that the 
probability $\Pi$ that a new vertex will be connected to vertex $i$ depends on the 
connectivity $k_i$ of that vertex, such that 
\begin{equation}
\Pi(k_i)=k_i/\sum_j k_j . 
\label{X}
\end{equation}

\begin{figure}
\resizebox{\textwidth}{!}{\includegraphics{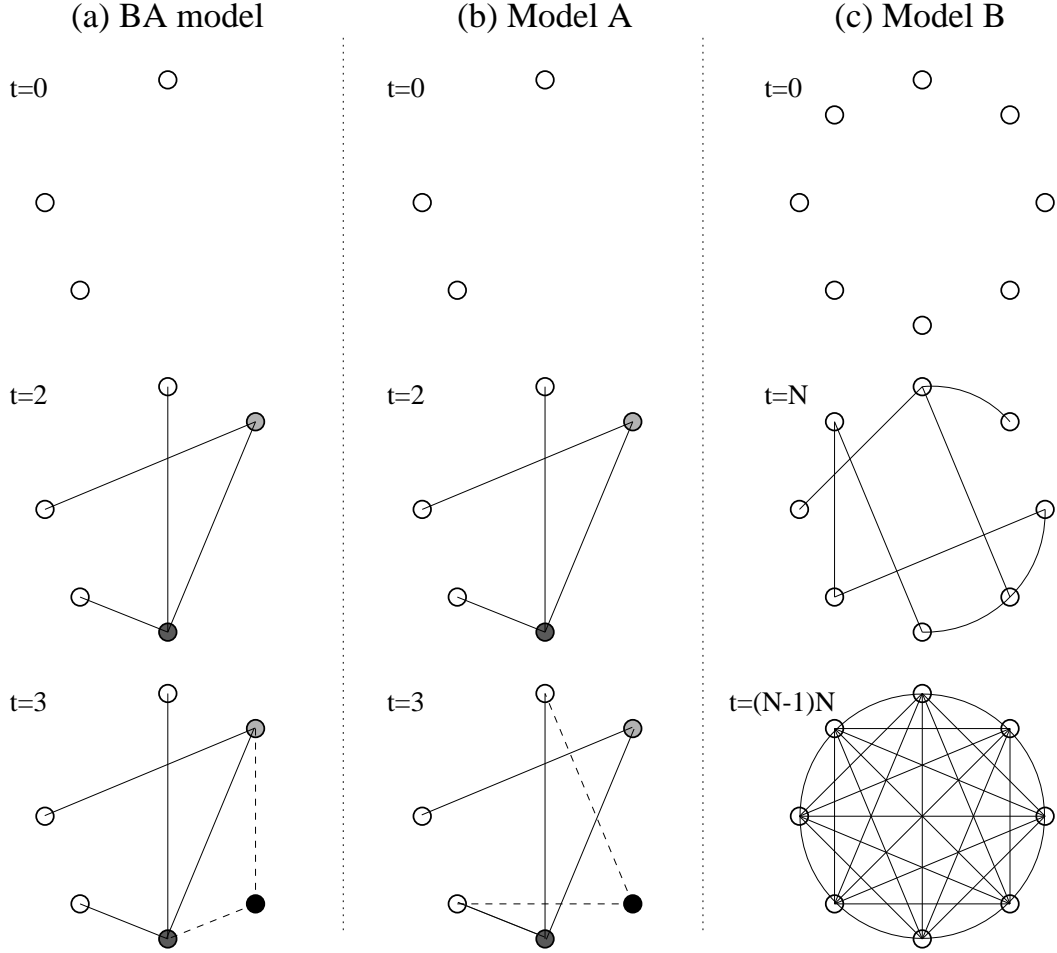}}
\caption{Illustration of the BA model and its variants. (a) BA model for $m_0=3$ and $m=2$. At $t=0$ the system consists of $m_0=3$ isolated vertices. At every timestep a new vertex is added, which is connected to $m=2$ vertices, preferentially to the vertices with high connectivity, determined by the rule (\ref{X}). Thus at $t=2$ there are $m_0+t=5$ vertices and $mt=4$ edges. At $t=3$ the sixth vertex is added, the two new edges being drawn with dashed lines. Due to preferential attachment the new vertex was linked to vertices with already high connectivity. (b) Model A with $m_0=3$ and $m=2$. At $t=0$ there are $m_0=3$ vertices and no edges. At every timestep a new vertex is added to the system, which is connected randomly to $m=2$ vertices already present. As in (a), at $t=2$ there are five vertices and four edges. At $t=3$ the sixth vertex is added to the system. The two new edges are drawn with dashed lines. Since preferential attachment is absent, the new vertex connects with equal probability to any vertex in the system. (c) Model B with $N=8$ vertices. In this model the number of vertices is fixed. At $t=0$ there are no edges. At every step a new edge is introduced, one end being added to a randomly selected vertex, the other end folowing preferential attachment (\ref{X}). At $t=N$ there are eight edges in the considered example, while at $t=N(N-1)/2$ the system is fully connected.}
\end{figure}

After $t$ timesteps the model leads to a random network 
with $N=t+m_0$ vertices and $mt$ edges. As Fig.$\,$4a shows, 
this network evolves into a scale-invariant state, the probability 
that a vertex has $k$ edges following a power-law with an 
exponent $\gamma_{model}=2.9\pm 0.1$. The 
scaling exponent is independent of $m$, the only 
parameter in the model. Since the power-law observed for real networks describes 
systems of rather different sizes at different stages of their 
development, one expects that a correct 
model should provide a distribution whose main features are 
independent of time. Indeed, as Fig.$\,$4b demonstrates, $P(k)$ is 
independent of time (and, subsequently, independent of the system 
size $N=m_0+t$), indicating that despite its continuous growth, 
the system organizes itself into a {\it scale-free stationary state}.

\begin{figure}[ht]
\rotatebox{270}{\resizebox{\textwidth}{!}{\includegraphics{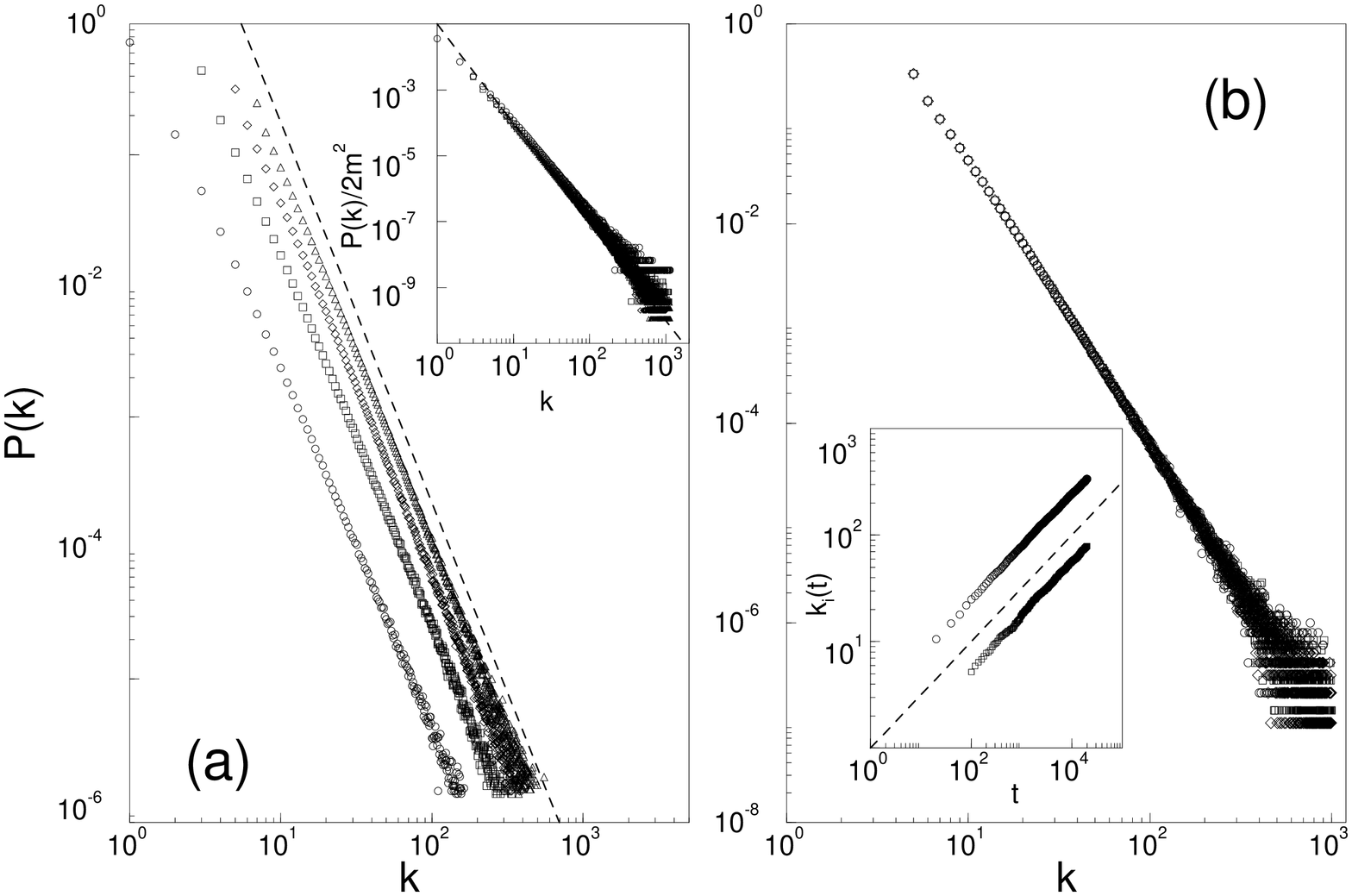}}}
\vspace{-3cm}
\caption{(a) Connectivity distribution of the BA model, with $N=m_0+t=300,000$ and $m_0=m=1$ (circles), $m_0=m=3$ (squares), $m_0=m=5$ (diamonds) and $
m_0=m=7$ (triangles). The slope of the dashed line is $\gamma=2.9$. The inset  shows the rescaled distribution (see text) $P(k)/2m^2$ for the same values of $m$, the slope of the dashed line being $\gamma=3$. (b) $P(k)$ for $m_0=m=5$ and system sizes $N=100,000$ (circles), $N=150,000$ (squares) and $N=200,000$ (diamonds). The inset shows the time-evolution for the connectivity of two vertices, 
added to the system at $t_1=5$ and $t_2=95$. Here $m_0=m=5$, and the dashed line has slope $0.5$, as predicted by Eq.$\,$(\ref{sqrt}).}
\end{figure}

We next describe a method to calculate analytically the probability $P(k)$,
allowing us to determine exactly the scaling exponent $\gamma$.
The combination of growth and preferential attachment leads
to an interesting dynamics of the individual vertex connectivities.
The vertices that have the most connections
are those that have been added at the early stages of the network
development, since vertices grow proportionally
to their connectedness relative to the rest of the vertices.
Thus some of the oldest vertices have a very
long time to acquire links, being responsible for
the high-$k$ part of $P(k)$.
The time dependence of the connectivity of a given vertex
can be calculated analytically using a mean-field approach.
We assume that $k$ is continuous, and thus the probability 
$\Pi(k_i)=k_i/\sum_j k_j$ can be interpreted as a continuous rate 
of change of $k_i$.
Consequently, we can write for a vertex $i$
\begin{equation}
\frac{\partial k_i}{\partial t} = A \Pi(k_i)=A \frac{k_i}{\sum_{j=1}^{m_0+t-1}k_j}.
\end{equation}
Taking into account that  $\sum_j k_j=2mt$ and the change in connectivities 
at a time step is $\Delta(k)=m$,
we obtain that $A=m$, leading to
\begin{equation}
\frac{\partial k_i}{\partial t}=\frac{k_i}{2t}. 
\end{equation}
The solution of this equation, with the initial condition 
that vertex $i$ was added to 
the system at time $t_i$ with connectivity $k_i(t_i)=m$, is 
\begin{equation}
k_i(t)=m\left(\frac {t}{t_i}\right)^{0.5}. 
\label{sqrt}
\end{equation}

As the inset of Fig.$\,$4b shows, the numerical results are in good agreement 
with this prediction. 
Thus older (smaller $t_i$) vertices increase their 
connectivity at the expense of the younger (larger $t_i$) vertices,  
leading with time to some vertices that are highly connected, 
a ``rich-gets-richer'' 
phenomenon that can be easily detected in real networks. 
Furthermore, this property can be used to calculate $\gamma$ 
analytically. Using (\ref{sqrt}), the probability that a vertex  
has a connectivity $k_i(t)$ smaller than $k$, $P(k_i(t)<k)$, 
can be written as 
\begin{equation}
P(k_i(t)<k)=P(t_i>{{m^2t}\over{k^2}}). 
\end{equation}
Assuming 
that we add the vertices at equal time intervals to the system, 
the probability density of $t_i$ is 
\begin{equation}
P_i(t_i) = \frac{1}{m_0+t}. 
\end{equation}
Substituting this into Eq. (4) we obtain that 
\begin{equation}
P(t_i>{{m^2t}\over{k^2}})=1-P(t_i\leq {{m^2t}\over{k^2}})=1-{{m^2t}\over{k^2(t+m_0)}}.
\end{equation} 
The probability density for $P(k)$ can be obtained using
\begin{equation}
P(k)=\frac{\partial P(k_i(t)<k)}{\partial k}=\frac{2m^2t}{m_0+t}\, \frac{1}{k^3}
\label{cube},
\end{equation} 
predicting 
\begin{equation}
\gamma=3,
\end{equation}
independent of $m$. 
Furthermore, Eq. (\ref{cube}) also predicts that the coefficient $A$ 
of the power-law distribution, $P(k) \sim A k^{-\gamma}$, is proportional 
to the square of the average
connectivity of the network, i.e., $A \sim m^2$. In the inset of Fig.$\,$4a 
we show $P(k)/2m^2$ vs. $k$. The curves obtained for different $m$ collapse
into a single one, supporting the analytical result (\ref{cube}).
  
\section{Limiting cases of the scale-free model}
\subsection{Model A}
The development of the power-law scaling in the scale-free model indicates 
that growth and preferential attachment play an important 
role in network development. To verify that both ingredients 
are necessary, we investigated two variants of the BA model.
The first variant, that we refer to as model A,
keeps the growing character of the network, but preferential 
attachment is eliminated.
The model is defined as follows (see Fig.$\,$3b):

(1) {\it Growth} : Starting with a small number of vertices ($m_0$),
at every time step we add a new vertex with $m(\le m_0)$ 
edges.

(2) {\it Uniform attachment} : We assume that the new vertex connects 
{\it with equal probability} to the vertices already present in the system,
i.e. $\Pi(k_i)=1/(m_0+t-1)$, independent of $k_i$.

Fig.$\,$5a shows the probability $P(k)$ obtained for different values of $m$, 
indicating
that in contrast with the scale-free model, $P(k)$
has an exponential form
\begin{equation}
P(k)=B \exp(-\beta k)
\label{exp}
\end{equation}

\begin{figure}[ht]
\rotatebox{270}{\resizebox{\textwidth}{!}{\includegraphics{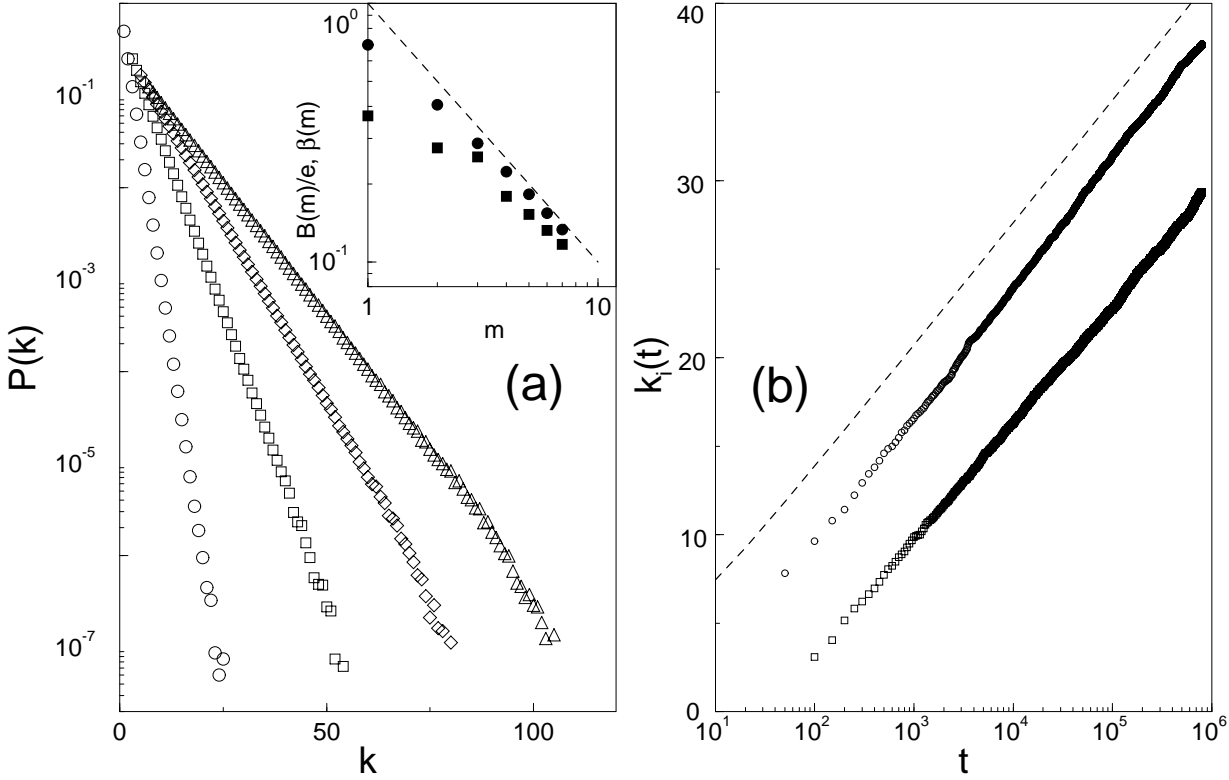}}}
\vspace{-3cm}
\caption{(a) The connectivity distribution for model A for $m_0=m=1$ (circles), $m_0=m=3$ (squares), $m_0=m=5$ (diamonds) and $m_0=m=7$ (triangles). The system size is $N=800,000$. The inset presents the scaling of the coefficients $B$ and $\beta$ (see Eq.$\,$(\ref{exp})) with $m$,
the dashed line following the prediction (18). (b) Time evolution for the connectivity of two vertices 
added to the system at $t_1=7$ and $t_2=97$. Here $m_0=m=3$. The dashed line follows $k_i(t)=m\ln(m_0+t-1)$ as predicted by (\ref{evol_a}).}
\end{figure}

We can use the mean field arguments developed in the previous section to 
calculate analytically the expression for $P(k)$. 
The rate of change of the connectivity of vertex $i$ in this case is given by
\begin{equation}
\frac{\partial k_i}{\partial t}=A\Pi(k_i)=\frac{A}{m_0+t-1}.
\end{equation}
At one timestep $\Delta(k)=m$, implying that $A=m$.
Solving the equation for $k_i$, and taking into account that $k_i(t_i)=m$, 
we obtain 
\begin{equation}
\label{evol_a}
k_i=m\left(\ln(m_0+t-1)-\ln(m_0+t_i-1)+1\right),
\end{equation}
a logarithmic increase with time, verified by the numerical simulations (see Fig.$\,$5b). 

The probability that vertex $i$ has connectivity $k_i(t)$ smaller than $k$  is 
\begin{equation}
P(k_i(t)<k)=P\left(t_i>(m_0+t-1)\exp(1-\frac k m)-m_0+1\right).
\label{diff}
\end{equation}
Assuming that we add the vertices
uniformly to the system, we obtain that 
\begin{eqnarray}
P\left(t_i>(m_0+t-1)\exp(1-\frac km)-m_0+1\right) \nonumber \\
= 1-\frac{(m_0+t-1)\exp(1-\frac km)-m_0+1}{m_0+t}. 
\end{eqnarray}
Using Eq. (10) and assuming long times, we obtain
\begin{equation}
P(k)=\frac{e}{m}\exp(-\frac{k}{m}) ,
\end{equation}
indicating that in (\ref{exp}) the coefficients are
\begin{equation}
B=\frac{e}{m},
~~ \beta =\frac{1}{m}.
\end{equation}
Consequently, the vertices in the model have the characteristic connectivity
\begin{equation}
k^* = {1 \over \beta} = m,
\end{equation}
which coincides with half of the average connectivities of the vertices
in the system, since $\left<k\right>=2m$.
As the inset of Fig.$\,$5a demonstrates the numerical results approach 
asymptotically the theoretical predictions. The exponential 
character of the distribution for this model indicates that the 
absence of preferential attachment eliminates the scale-free feature of the
BA model. 

\subsection{Model B}
This model tests the hypothesis that the growing character of the model 
is essential to sustain the scale-free state observed in the real systems.
Model B is defined as follows (see Fig.$\,$3c):

We start with $N$ vertices and no edges. At each time step 
we randomly select a vertex and connect it with 
probability $\Pi(k_i)=k_i/\sum_j k_j$ to vertex $i$ in the system. 

Consequently, in comparison with the BA model, this variant eliminates
the growth process, the numbers of vertices staying  constant
during the network evolution.
While at early times the model exhibits power-law scaling (see Fig.$\,$6,
$P(k)$ is not stationary: Since $N$ is constant, and the number 
of edges increases with time, after $T\simeq N^2$ 
timesteps the 
system reaches a  state in which all vertices are connected. 

The time-evolution of the individual connectivities can be calculated analytically using the mean field approximation developed for the previous models. The rate of change of the connectivity of vertex $i$ has 
two contributions: the first describes the probability that the 
vertex is chosen randomly as the origin of the 
link, $\Pi_{random}(k_i)=1/N$ and the second is proportional 
to $\Pi(k_i)=k_i/\sum_j k_j$, describing the probability 
that an edge originating from a randomly selected vertex 
is linked to vertex $i$:
\begin{equation}
\frac{\partial k_i}{\partial t} =A \frac{k_i}{\sum_{j=1}^{N}k_j}+\frac 1N. 
\end{equation}

Taking into account that $\sum_j k_j=2t$ and that the change in connectivities during 
one timestep is $\Delta(k)=2$, and excluding from the summation 
edges originating and terminating in the same vertex, we obtain $A=N/(N-1)$, 
leading to
\begin{equation}
\frac{\partial k_i}{\partial t}=\frac{N}{N-1}\,\frac{k_i}{2t}+\frac{1}{N}. 
\end{equation}
The solution of this equation has the form
\begin{equation}
\label{evol}
k_i(t)=\frac{2(N-1)}{N(N-2)}t+Ct^{\frac{N}{2(N-1)}}.
\end{equation}
Since $N >> 1$, we can approximate $k_i$ with
\begin{equation}
k_i(t)=\frac{2}{N} t + C t^{1/2}
\end{equation}
Since the number of vertices is constant, we do not 
have ``introduction times'' $t_i$ for the vertices. There exists, 
however, a time time analogous 
to $t_i$: the time when vertex $i$ was selected for the first time
as the origin of an 
edge, and consequently its connectivity changed from $0$ to $1$. 
Equation (\ref{evol}) is valid only for $t>t_i$, and all
vertices will follow this dynamics only after $t\geq N$.
The constant $C$ can be determined from the condition that $\sum_j k_j=2t$, and has the value 
\begin{equation}
C=0,
\end{equation}
thus 
\begin{equation}
k_i(t)\simeq\frac{2}{N}t.
\label{evol_1}
\end{equation}
The numerical results shown in Fig.$\,$6b agree well with this prediction,
indicating that after a transient time of duration $t\simeq N$ the 
connectivity increases linearly with time.

\begin{figure}[ht]
\rotatebox{270}{\resizebox{\textwidth}{!}{\includegraphics{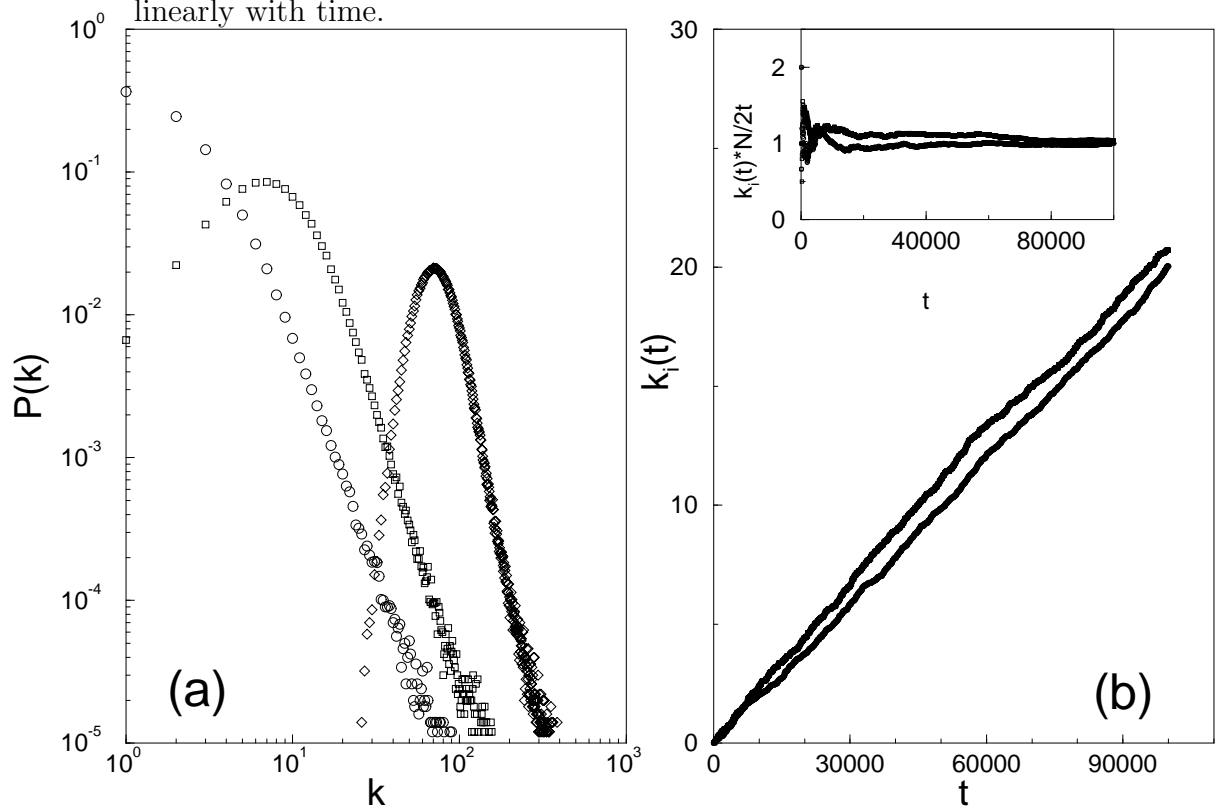}}}
\vspace{-3cm}
\caption{(a) The connectivity distribution for model B for $N=10,000$ and $t=N$ (circles), $t=5N$ (squares), and $t=40N$ (diamonds) and $t=40N$. (b) Time dependence of the connectivities of two vertices. The system size is $N=10,000$. The inset shows the connectivities rescaled by $N/2t$, supporting the theoretical prediction $k_i(t)N/2t\rightarrow 1$.}
\end{figure}
  
Since the mean-field approximation used above predicts that after a transient period the connectivities of all vertices should have the same value given by Eq. (\ref{evol_1}), we expect that the connectivity distribution becomes a Gaussian around its mean
value. Indeed, Fig.$\,$6a illustrates that as time increases, the shape of $P(k)$ changes from the initial power-law to a Gaussian.

The failure of models A and B  in leading to a scale-free distribution 
indicates that both ingredients, 
namely growth and preferential attachment, are needed 
to reproduce the stationary power-law distribution observed in
real networks.

\section{Discussion and conclusions}
In the following we discuss some of the immediate extensions of the present
work.

(i) A major assumption in the model was the use of a linear relationship
between $\Pi(k_i)$ and $k_i$, given by (\ref{X}).
However, at this point there is nothing to guarantee us that $\Pi(k)$ is linear, i.e. in general we could assume that $\Pi(k) \sim k^{\alpha}$, where $\alpha \neq 1$.
The precise form of $\Pi(k)$ could be determined numerically
by comparing the topology of real networks at not too distant times.
In the absence of such data, the linear relationship seems to be the most 
efficient way to go. In principle, if nonlinearities are present
(i.e. $\alpha \neq 1$), that could affect the nature of the power-law
scaling. This problem will be addressed in future work \cite{rha}.

(ii) An another quantity that could be tested explicitelly
is the time evolution of connectivities in real networks. For the scale-free
model we obtained that the connectivity increases as a power of time
(See Eq. (\ref{sqrt})).
For model A we found logarithmic time dependence (Eq. (\ref{evol_a})), while for model B linear (Eq. (\ref{evol_1})).
Furthermore, if we introduce $p_{\mathrm {ER}}=at$ in the ER model, one can easily show that
$\langle k\rangle_{\mathrm {ER}}(t) \sim t$. If time resolved data on network connectivity becomes
available, these predictions could be explicitelly tested for real networks,
allowing us to distinguish between the different growth mechanisms.

(iii) In the model we assumed that new links appear only when new vertices are 
added to the system. In many real systems, including the movie actor 
networks or
the www, links are added continuously. Our model can be easily extended to
incorporate the addition of new edges.
Naturally, if we add too many edges, the system becomes fully connected.
However, in most systems the addition of new vertices (and the growth
of the system) competes with the addition of new internal links.
As long as the growth rate is large enough, we believe that the system
will remain in the universality class of the BA model, and
will continue to display scale-free features.

(iv)  Naturally, in some systems we might witness the reconnection or rewiring 
of the
existing links. Thus some links, that were added when a new vertex was
added to the system, will break and reconnect with other vertices,
probably still obeying preferential attachment. If 
reattachment dominates over growth (i.e. addition of new links by new 
vertices), the system will undergo a process similar to ripening:
the very connected sites will acquire all links. This will 
destroy the power-law scaling in the system. However,
similarly to case (iii) above, as long as the growth process dominates
the dynamics of the system, we expect that the scale-free state will prevail.

(v) The above discussion indicates that there are a number of ``end-states''
or absorbing states for random networks, that include the scale-free state, when
power-law scaling prevails at all times, the fully connected state,
which will be the absorbing state of the ER model for large $p$, and
the ripened state, which will characterize the system described in point (iv).
Note that the end state of the WS model, obtained for $p_{\mathrm {WS}}=1$, is the
ER model for $p_{\mathrm {ER}}=z/N$.
The precise nature of the transition between these states is still an
open question, and will be the subject of future studies \cite{rha}.

(vi) Finally, the concept of universality classes has not been properly
explored yet in the context of random network models. For this we
have to define scaling exponents that can be measured for {\it all} random
networks, whether they are generated by a model or a natural process.
The clustering of these exponents for different systems might 
indicate that there are a few generic universality classes 
characterizing complex networks. Such studies have the potential to lead
to a better
understanding of the nature and growth of random networks in general.

Growth and preferential attachment are mechanisms common to a number of 
complex systems, including business networks \cite{arthur}, 
social networks (describing individuals  or organizations), 
transportation networks \cite{banavar}, etc. Consequently, we 
expect that the scale-invariant state,  observed in all systems for 
which detailed data  has been available to us, is a generic 
property of many complex networks, its applicability reaching  
far beyond the quoted examples. A better description  of these 
systems would help in understanding other complex systems as well, for 
which so far less topological information is available, including  
such important examples as genetic or signaling networks in 
biological systems. 
Similar mechanisms could explain the origin of the 
social and economic disparities governing competitive systems, 
since the scale-free inhomogeneities are the inevitable  
consequence  of self-organization due to the local decisions made by the 
individual vertices, based on information that is biased 
towards the more visible (richer) vertices, irrespective of the 
nature and the origin of this visibility.

\ack{
This work was supported by the NSF Career Award DMR-9710998.
}


\begin{thebibliography}{999}
\bibitem{note}
Note that in addition to the distribution of incoming links, 
the www displays a number of other scale-free features, 
characterizing the organization of the webpages within a 
domain \cite{huberman_web}, the distribution of 
searches \cite{huberman_science}, 
or the number of links per webpage \cite{diam}.
\end{thebibliography}
\end{document}